\documentclass[preprintnumbers,prd,twocolumn,showpacs,showkeys,floatfix,preprintnumbers,letterpaper,amsmath,amssymb,superscriptaddress]{revtex4-1}
\usepackage{graphicx}
\usepackage{color}

\newcommand{\bea}{\begin{eqnarray}}
\newcommand{\eea}{\end{eqnarray}}
\newcommand{\nn}{\nonumber}
\newcommand{\be}{\begin{equation}}
\newcommand{\ee}{\end{equation}}
\newcommand{\Mpl}{M_{\rm Pl}}
\newcommand{\Lam}{\Lambda}

\begin{document}

\title{Extreme parameter sensitivity in quasidilaton massive gravity}

\author{Stefano Anselmi}
\email{stefano.anselmi@case.edu}
\affiliation{Department of Physics, Case Western Reserve University, Cleveland, OH 44106-7079 -- USA}

\author{Diana L{\' o}pez Nacir}
\email{dlopez\_n@ictp.it}
\affiliation{The Abdus Salam International Center for Theoretical Physics, Strada Costiera 11, I-34151 Trieste -- Italy}
\affiliation{Departamento de F\'\i sica and IFIBA, FCEyN UBA, Facultad de Ciencias Exactas y Naturales, Ciudad Universitaria, Pabell\' on I, 1428 Buenos Aires  --  Argentina}

\author{Glenn D. Starkman}
\email{glenn.starkman@case.edu}
\affiliation{Department of Physics, Case Western Reserve University, Cleveland, OH 44106-7079 -- USA}

\date{\today}

\begin{abstract}
We reanalyze the behavior of Friedmann-Lema\^\i tre-Robertson-Walker  cosmologies in the recently proposed quasidilaton massive-gravity model,
and discover that the background dynamics present hitherto unreported features that require unexpected fine-tuning of the additional fundamental parameters of the theory
for an observationally consistent background cosmology. We also identify new allowed regions in the parameters space and exclude some of the previously considered ones.
The evolution of the mass of gravitational waves reveals non-trivial behavior, exhibiting a mass-squared that may be negative in the past,
and that presently, while positive, is larger than the square of the Hubble parameter, $H_0^{2}$. 
These properties of the gravity-wave mass have the potential to lead to observational tests of the theory. 
While quasidilaton massive gravity is known to have
issues with stability at short distances, the current analysis is  a first step toward the investigation of the more stable extended quasidilaton massive gravity theory,
with some expectation that both the fine-tuning of parameters and the interesting behavior of the gravity-wave mass will persist.
\end{abstract}


\maketitle

\section{\label{Intro}Introduction} The standard cosmological model,   $\Lambda$CDM, describes the acceleration of the universe by properly adjusting
the  cosmological constant $\Lambda$. While this  simple model is consistent with current observational data,  other models provide
alternative explanations of this acceleration. For example, some models attribute the acceleration to the presence of a dynamical component known as  dark energy \cite{Creminelli:2008wc,1999PhRvL..82..896Z,2000PhRvD..62b3511C},
and others to a modification of the gravitational laws on cosmological distances \cite{2004JHEP...05..074H,Nojiri:2006ri,2010RvMP...82..451S,2000PhLB..485..208D,2004LRR.....7....7M}.  The questions will be to what extent  it is possible to discriminate among the different models from observations, and whether any of the models are better at fitting the data than what is currently the most parsimonious explanation, $\Lambda$.

The next generation of  experiments (such as EUCLID \cite{euclid}
or DESI \cite{desi}) will provide an unprecedented amount of observational data. However, there  is now a wide range of 
candidate theories.  For instance, different modifications of general relativity  primarily in the infrared have been considered by many authors (see \cite{Clifton:2011jh} for a recent review), and probably still  others have yet to be proposed. Ultimately, the predictions of each candidate model must be confronted with data. This includes not just cosmological data but data on all scales where the models make calculable predictions that can be tested observationally or experimentally.

Within one interesting class of theories, the current acceleration era is associated to the presence of a mass term for the graviton (for a historical overview,
motivations and an updated description of different proposed massive gravity theories, see  \cite{2014LRR....17....7D,Tolley:2015oxa}). 
Here we consider a particular modification of general relativity 
known as quasidilaton massive gravity (QDMG), 
which we summarize in the Section \ref{QD}. This theory was proposed in \cite{2013PhRvD..87f4037D}, as an extension of the  dRGT theory of massive gravity \cite{2010PhRvD..82d4020D, 2011PhRvL.106w1101D}, and contains an additional scalar degree of freedom: the quasidilaton.
The main motivation for such extension is the absence of isotropic and homogeneous cosmological
background solutions in  dRGT  \cite{PhysRevD.84.124046}.
Indeed, it has been shown that  QDMG has solutions with spatially flat Friedmann-Lema\^\i tre-Robertson-Walker  (FLRW) background
metrics \cite{2013PhRvD..87f4037D}. Moreover, it has been found
that (even in the absence of a cosmological constant) there are solutions for which at late times the metric approaches to a de Sitter metric, providing a plausible
(self-accelerating) explanation of the accelerated expansion of the universe \cite{2013PhRvD..87f4037D, 2013PhRvD..87l3536G, 2014PhRvD..89h3518B}. The quasidilaton 
theory has three  parameters more than $\Lambda$CDM. 

In this paper we will perform a careful analysis of the background cosmological evolution, taking into account  the main goal of describing the observed expansion history of the universe.
While other authors have made preliminary investigations \cite{2013PhRvD..87l3536G,2014PhRvD..89h3518B} of the background evolution in QDMG, a more detailed reexamination reveals important new insights. The allowed set of parameters split into two disconnected regions characterized by ``low'' and ``high'' values of a dimensionless parameter of the theory, $\omega$  (which multiplies the kinetic term of the quasidilaton).  In the region with low values of $\omega$,  while viable background solutions exist for a wide range of values of the  Lagrangian parameter nominally called the graviton mass $m_g$,  with  $m_g \sim {\cal{O}}(H_0)$,  a careful fine-tuning of the dimensionless constants $\alpha_3$ and $\alpha_4$ is required. The permitted values of $\alpha_3$, $\alpha_4$ and $m_g$ thus describe a very thin 2-dim surface $m_g(\alpha_3, \alpha_4)$ in the $\{\alpha_3, \alpha_4, m_g\}$ parameter space. In the other region, the parameter $m_g$   is  constrained to be much smaller than $H_0$, and the larger it is, the narrower the 2-dim surface of allowed $\alpha_3$ and $\alpha_4$. 

The paper is organized as follows. After  summarizing the theory QDMG in Section \ref{QD},  
in Section \ref{DE} we present the dynamical equations. 
In Section \ref{FP} we analyze the existence of viable  de Sitter fixed-point attractors.  
By exploring the 4-dimensional parameter space of the theory,  in Section \ref{PF}, 
we assess the viability of a self-accelerating explanation  of the current expansion of the universe.
An important outcome of our analysis is that,  
in order to reproduce an expansion history consistent with  data, the graviton mass parameter must also be fine-tuned to a value that depends on other parameters of the model.

In Section \ref{GW} we study the evolution of the mass of 
gravitational waves $M_{GW}$ for the allowed set of parameters.  
We find the current value of $M_{GW}$ to be generically larger than the current Hubble constant $H_0$
even when we set the graviton mass parameter $m_g  \ll H_0$.
 In the past (for example at redshifts relevant for the Cosmic Microwave Background) 
 $M_{GW}$ can be either real or  imaginary. 
 For a conservative choice  of $0<m_g\leq H_0$, $\vert M_{GW}(t) \vert <H(t) $ in the past, with
 $\vert M_{GW}\vert\lesssim 10^{-2} H$ at last scattering.
 While this precludes the development of a catastrophic instability when $M_{GW}$ is imaginary, 
nevertheless potentially there could   be observable cosmological signatures.  These merit further investigation \cite{PhysRevD.81.023523, 2012CQGra..29w5026E, 2009JCAP...08..033B}.

\section{\label{QD} Theory of quasidilaton massive gravity}
We consider the action for the quasidilaton theory \cite{2013PhRvD..87f4037D}: 
\begin{eqnarray}
S & = & S_{EH}+S_{\sigma}\nonumber\\
&=& \frac{M_{{\rm Pl}}^{2}}{2}\int d^{4}x\sqrt{-g}\biggl[R-\frac{\omega}{M_{{\rm Pl}}^{2}}\partial_{\mu}\sigma\partial^{\mu}\sigma  \\
&&\quad\quad + 2m_{g}^{2}(\mathcal{L}_{2}+\alpha_{3}\mathcal{L}_{3}+\alpha_{4}\mathcal{L}_{4})\biggr],\nonumber\label{eq:action}
\end{eqnarray}
where $M_{\rm Pl}$ is the Planck mass and, in addition to the Einstein-Hilbert action $S_{EH}$, 
a contribution $S_{\sigma}$ characterizes the quasidilaton  scalar field $\sigma$.  In addition 
to the quasidilaton kinetic term, $S_\sigma$ includes three interaction terms:
Here
\begin{eqnarray}
&&\mathcal{L}_{2}  \equiv  \frac{1}{2}\,([\mathcal{K}]^{2}-[\mathcal{K}^{2}])\,,\\
&&\mathcal{L}_{3}  \equiv  \frac{1}{6}\,([\mathcal{K}]^{3}-3[\mathcal{K}][\mathcal{K}^{2}]+2[\mathcal{K}^{3}])\,,\\
&&\mathcal{L}_{4}  \equiv  \frac{1}{24}\,([\mathcal{K}]^{4}-6[\mathcal{K}]^{2}[\mathcal{K}^{2}]+3[\mathcal{K}^{2}]^{2}+ \nn \\
&&\qquad\,+\,8[\mathcal{K}][\mathcal{K}^{3}]-6[\mathcal{K}^{4}])\,,
\end{eqnarray}
 with  square brackets denoting a trace, and
\begin{eqnarray}
\mathcal{K}_{\ \nu}^{\mu}&\equiv&\delta_{\ \nu}^{\mu}-e^{\sigma/M_{{\rm Pl}}}\left(\sqrt{g^{-1}{f}}\right)_{\ \ \nu}^{\mu}\,.
\end{eqnarray}
The non-dynanmical ``fiducial metric'' is built from four St\"uckelberg fields $\phi^{a}$ ($a=0,\cdots,3$), 
\begin{eqnarray}
{f}_{\mu\nu}&\equiv&\eta_{ab}\partial_{\mu}\phi^{a}\partial_{\nu}\phi^{b}.
\end{eqnarray}

In the space of
St\"uckelberg fields, the theory  enjoys the Poincare symmetry \cite{2013PhRvD..87f4037D}
\begin{equation}
\phi^{a}\to\phi^{a}+c^{a}\,,\qquad\phi^{a}\to\Lambda_{b}^{a}\phi^{b}\,,
\end{equation} and in addition, there is a  global symmetry given by
\begin{equation}
\sigma\to\sigma+\sigma_{0}\,,\qquad\phi^{a}\to e^{-\sigma_{0}/M_{{\rm Pl}}}\,\phi^{a}\,,\label{eqn:symmetry}
\end{equation}  with $\sigma_{0}$  an arbitrary constant. 

The addition of $S_\sigma$ to the action, introduces four new parameters:
the dimensionless kinetic coupling $\omega$, the graviton mass parameter $m_{g}$,
and the coupling constants
$\alpha_{3}$ and $\alpha_{4}$. As shown below, the cosmological solution
depends sensitively on the values of these parameters.

\section{The background cosmological equations}
\label{DE}

We  consider  a spatially flat 
Friedmann-Lema\^\i tre-Robertson-Walker ansatz, for which
\begin{eqnarray}
ds^{2} & = & -N(t)^{2}dt^{2}+a(t)^{2}\delta_{ij}dx^{i}dx^{j}\,,\\
\phi^{0} & = & \phi^{0}(t)\,,\\
\phi^{i} & = & x^{i}\,,\\
\sigma & = & \bar{\sigma}(t)\,.
\end{eqnarray}
The  fiducial metric $f_{\mu\nu}$ reduces to 
\begin{equation}
{f}_{00}=-n(t)^2\,, \quad {f}_{ij}=\delta_{ij}\,,
\end{equation}
where
\begin{equation}
\label{defn}
 n(t)^2 \equiv \bigl(\dot{\phi}^{0}\bigr)^{2}\,.
\end{equation}
The minisuperspace action for the background metric and fields can now be written as
\begin{eqnarray}\label{eq:actionB}
S/V  &=& M_{{\rm Pl}}^{2}\int dt
	\left\{-3 \frac{a^3}{N}\left(\frac{\dot{a}}{a}\right)^2
+ a^3\frac{w}{M_{\rm Pl}}\frac{\dot{\sigma}^2}{2N}\right.\\
&&\quad \left. + N a^3 m_{g}^{2}(\mathcal{L}_{2}+\alpha_{3}\mathcal{L}_{3}+\alpha_{4}\mathcal{L}_{4})\right\},\nonumber
\end{eqnarray}
where
\begin{eqnarray}
\mathcal{L}_{2} &= &3(X-1)(-2+X(1+r))\,,\\
\mathcal{L}_{3} & =&-(X-1)^2(-4+X(1+3r))\,,\\
\mathcal{L}_{4} & = &(X-1)^3(-1+rX)\,,
\end{eqnarray}
and we have defined
\begin{eqnarray}
X & \equiv & \frac{e^{\bar{\sigma}/M_{\rm Pl}}}{a}\,,\\
r & \equiv & \frac{n}{N}\, a\,.
\end{eqnarray}

Varying the action with respect to $\phi^0(t)$ leads to
\begin{equation}
\label{Constraint}
\partial_t\left[ \frac{\dot{\phi^0}}{n}a^4\, G_2(X)\right] = \partial_t\left[a^4\, G_2(X)\right] =0\,, 
\end{equation}
where  $G_2(X)=X(1-X)J(X)$, with
\begin{equation}
 J(X) \equiv 3+3(1-X)\alpha_{3}+(1-X)^{2}\alpha_{4}.
\end{equation}
We  use time reparameterization freedom to set $N=1$.

In summary, the independent background equations are: 
\begin{itemize}
\item the constraint equation (\ref{Constraint}), or its integral
\begin{equation}
G_2(X) = \frac{C}{a^{4}}.
\label{int_const}
\end{equation}
\item the  Friedman equation,
\begin{equation}
\label{Friedman}
3H^2=\frac{\omega}{2}\left(\frac{\dot\sigma}{M_{\rm Pl}}\right)^2
	+ 3m_g^2  G_1(X)+\frac{\rho_m+\rho_r}{M_{\rm Pl}^2},
\end{equation} 
\noindent
where 
\begin{equation}
G_1(X)\equiv \frac{X\!\!-\!\!1}{3}  \left[\alpha_3 (X\!\!-\!\!1)^2 - 3(X\!\!-\!\!1)+ J(X)\right]
\end{equation}
and we have included the contributions of matter and radiation;  
\item the conservation of the stress-energy tensor obtained from  $S_{\sigma}$
 \begin{equation}
(\ddot{\sigma}+3H\dot{\sigma})\omega\dot{\sigma}+3 M_{\rm Pl} m_g^2(\dot{\sigma}-r H M_{\rm Pl})XG_1'(X)=0,
\label{conservSigma}
\end{equation} 
where a prime means derivative with respect to  $X$; 
\item the conservation of the stress-energy tensors of matter, $\dot{\rho}_m=-3H{\rho_m}$ and of radiation  
$\dot{\rho}_{r}=4H{\rho_r}$.
\end{itemize}
(Note that using the constraint equation (\ref{Constraint}) one can show that the equation obtained by taking the variation of $S_{\sigma}$ with respect to $\sigma$ is not an independent equation.)

\section{\label{FP} De Sitter fixed point analysis}

We start by investigating the future background evolution of the quasidilaton massive gravity model. 
The $\Lambda$CDM concordance model predicts the universe will approach a de Sitter phase in the future.
Though we do not know the future of the universe, 
we require our model to reproduce this prediction, 
consistent with recent practice \cite{2013PhRvD..87l3536G, 2014PhRvD..89h3518B}.

We rewrite eq.~(\ref{Friedman}) in terms of the relative energy densities
\begin{equation}\label{Friedman2}
1=\Omega_{DE}+\Omega_m+ \Omega_r,
\end{equation}
where
\be
\Omega_{DE}=\Omega_{\Lambda}+\Omega_{\sigma}\,,
\label{omega_DE_eq}
\ee
and
\begin{eqnarray}
\Omega_{m}&=&\frac{\rho_{m}}{3M_{\rm Pl}^2H^2}\,, \\
\Omega_{r}&=&\frac{\rho_{r}}{3M_{\rm Pl}^2H^2}\,, \\
 \Omega_{\Lambda}&=&\frac{m_g^2}{H^2}G_1(X),\label{OmLdef}\\
 \Omega_{\sigma}&=&\frac{\omega}{6H^2}\left(\frac{\dot{\sigma}}{M_{\rm Pl}}\right)^2. \label{Omsigma}
\end{eqnarray}
Employing eq. (\ref{Constraint}) and assuming that $X \neq 0$ we obtain
\be
\dot{\sigma}=\Mpl H	\left(1-\frac{4 G_{2}(X)}{X G_{2}'(X)} \right).\,
\label{sigmadot}
\ee
Eq. (\ref{int_const}) implies that as $a \to \infty$, $X \to \text{constant}$. Therefore the set of variables $\{\Omega_{m}$, $\Omega_{r}$, $X$, $\Omega_{\Lam}\}$ 
will approach constants in the asymptotic future. 
We study the dynamical stability of the system by means of the following equations:
\begin{eqnarray}
\frac{d\Omega_{r}}{dN}&=&-2 \Omega_{r}\left(2+\frac{\dot{H}}{H^{2}}\right) \label{au_1}\, , \\
\frac{d\Omega_{\Lambda}}{dN}&=&-2 \Omega_{\Lambda}\left(2\frac{G_{2}G_{1}'}{G_{1}G_{2}'}+\frac{\dot{H}}{H^{2}}\right) \label{au_2}\, , \\
\frac{dX}{dN}&=&-4 \frac{G_{2}}{G_{2}'} \label{au_3}\, . \\ \nonumber
\end{eqnarray}
$\frac{\dot{H}}{H^2}$ can be obtained by differentiating the first Friedmann equation (\ref{Friedman}) to obtain
 \begin{align}
\frac{\dot
H}{H^2}=\frac{9\Omega_m \!+\! 12\Omega_r \!+\! 12\frac{G_2}{G_2'}\Bigl[\frac{G_1'}{G_1
}\Omega_\Lambda
\!+\!\frac{\omega}{6}\frac{\rm d}{{\rm d}\xi}(1\!-\!4\frac{G_2}{\xi
G_2'})^2\Bigr]}{\omega \Bigl[1-4\frac{G_2}{\xi G_2'}\Bigr]^2-6}.
\label{HdotH2}
\end{align}

As noted above, we focus on de Sitter fixed points, and require that these critical points are attractors. The de Sitter critical points relative to the system 
(\ref{au_1}),  (\ref{au_2}) and  (\ref{au_3}) are given in Table \ref{stability1}, where 
\be
X_{\pm} \equiv 1+\frac{3}{2}\frac{\alpha_3}{\alpha_4}\pm\sqrt{\frac{9\alpha_3^2}{4\alpha_4^2}-\frac{3}{\alpha_4}}.
\ee

\begin{table*}[ht]
\begin{center}
\begin{tabular}{|c|c|c|c|c|c|c|c|c|c|c|c|}
\hline F. P.&
$\Omega_r$ & $\Omega_\Lambda$ & $X$ & $\Omega_m$ &  $\Omega_\sigma$ &Existence& Stability & Eigenvalues\\
\hline
 $A$&0 & $1-\omega/6$  & $X_+$ & 0 & $\omega/6$ &$0<\omega<6$,  $0\leq X_+$, $X_+\in {\cal{R}}$& Attractor  & -4,-4,-3 \\
\hline
 $B$&0 & $1-\omega/6$  & $X_-$ & 0 & $\omega/6$ &$0<\omega<6$,  $0\leq X_-$, $X_-\in {\cal{R}}$& Attractor  & -4,-4,-3 \\
\hline
C&0 & $1-3\omega/2$  & 0 & 0 & $3\omega/2$ & $0<\omega<2/3$&  Attractor  & -4,-4,-3 \\
\hline
\end{tabular}
\end{center}
\caption[crit]{\label{stability1} De Sitter fixed points.}
\end{table*}

To assess the stability we compute the  matrix form of the perturbation equations linearized  around each of the fixed points. Then, the linear asymptotic stability of each fixed point can be studied by analyzing the signs of the eigenvalues of that matrix. If the sign of the real part of every eigenvalues is negative, then the critical point is an attractor. 
The results are shown in the  last column of Table \ref{stability1}. 
 
This analysis indicates there are three possible late-time de Sitter fixed points,  $A$, $B$ and $C$. 
For each, constraints on the parameters $\alpha_{3}$ and $\alpha_{4}$  are obtained by requiring that 
$H^{2}>0$ and $X \geq 0$. 
For the point $A$ we obtain $\alpha_{3}>0$ and $0< \alpha_{4}<2\alpha_{3}^{2}/3$; 
for  $B$, $\alpha_{3}<-3$ and $-3-3\alpha_{3}\leq \alpha_{4}\leq 2\alpha_{3}^{2}/3$; 
for  $C$, $\alpha_{4}<-6-4\alpha_{3}$. 
Noticing that one must insist that $X \geq 0$, we found different constraints 
than \cite{2013PhRvD..87l3536G, 2014PhRvD..89h3518B}.

Consider more closely the fixed point B. 
Given the $\alpha_{3}$ and $\alpha_{4}$ constraints for $B$, we obtain $0<X_{-}<X_{+}<1$. 
The constraint equation (\ref{int_const}) implies that in the asymptotic future $G_{2}(X)=0$. 
Moreover (\ref{int_const}) requires that $G_{2}$ should be unbounded either above or below 
in order to have a past history. 
$G_{2}$ is a polynomial in X, so this is impossible if $0<X_{-}<1$ \footnote{Notice that $X=0$ and $X=1$ cannot be crossed in the past history.} as it is. Thus $B$ cannot be a well-defined fixed point. Recalling that $X \geq 0$, a similar argument can be applied to the point $C$. 

The allowed $\{\alpha_{3}, \alpha_{4}\}$ parameter region for $A$ entails that $1<X_{-}<X_{+}$. Therefore $A$ is the only well-defined de Sitter fixed point for the QDMG theory. We emphasize that our findings now differ from those of \cite{2013PhRvD..87l3536G, 2014PhRvD..89h3518B}, in that we exclude the points $B, C$.

\section{\label{PF}Cosmological evolution and parameter fixing}

\begin{figure}[t]
\begin{center}
\includegraphics[width=0.5\textwidth]{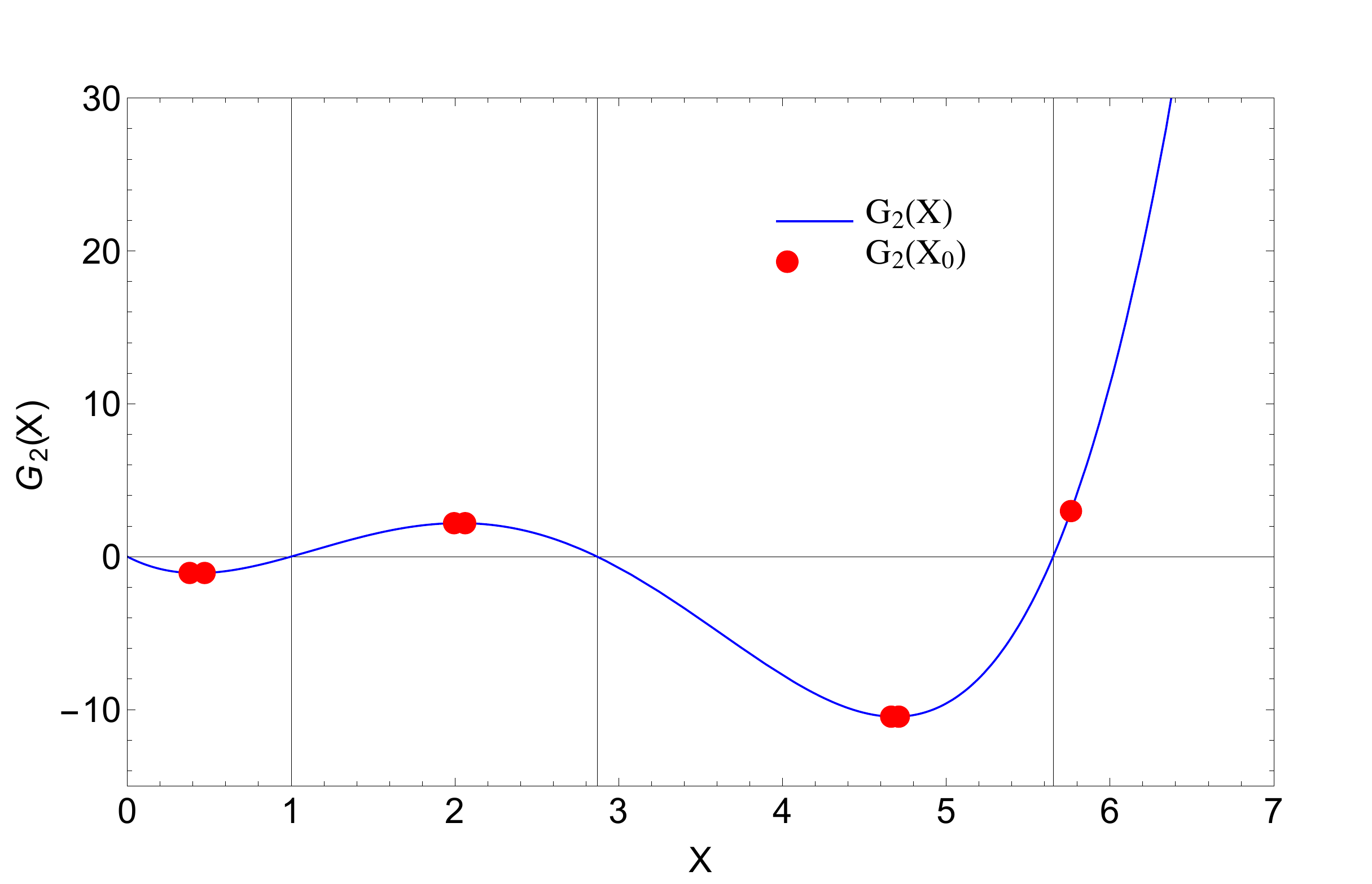}
\caption{The blue line shows the $G_{2}(X)$ function, while the red points are the $G_{2}(X_{0})=C$ values for $\{m_{g}=0.4H_{0},\, \omega=0.01,\, \alpha_{3}=0.75,\, \alpha_{4}=0.345\}$.}
\label{plotG2}
\end{center}
\end{figure}

The aim of this section is to study the evolution of the relevant background quantities in agreement with the results of the previous section and with the observed cosmological history. That depends on the initial conditions, on the expansion history and on the fixed point $A$. By means of this analysis we constrain the four parameters of quasidilaton massive gravity:  $\{m_{g}, \omega, \alpha_{3}, \alpha_{4}\}$. 

Given that we are dealing with the background energy density evolution, we can consider neutrinos to be relativistic, since the value of $\Omega_{r}$ is not negligible only in the radiation era when neutrinos were indeed relativistic. The spectrum of the CMB today is precisely measured, 
so we accurately determine $\Omega_{\gamma,0}$. For relativistic neutrinos, $\Omega_{\nu,0}$ is proportional to $\Omega_{\gamma,0}$.  Therefore we assume $\Omega_{r,0}$ is known and we fix it by $\Omega_{r,0}=\Omega_{\gamma,0}+\Omega_{\nu,0}=0.0000851$. 

To be definite we also fix $\Omega_{DE, 0}=0.72$, close to the best fit value \cite{2015arXiv150201589P}. That corresponds in $\Lambda$CDM to $z_{eq} \approx 3300$. We stress that this choice will not qualitatively affect our conclusions on the quasidilaton massive gravity background evolution. 

To fix the initial conditions we require $\Omega_{DE,0}=\Omega_{\sigma,0}+\Omega_{\Lambda,0}$, where $\Omega_{\sigma,0}$ and $\Omega_{\Lambda,0}$ are given by (\ref{OmLdef}), (\ref{Omsigma}). 
In this way we obtain a 9-th order polynomial that has no analytical solutions. 
In Fig.~\ref{plotG2} we plot the $G_{2}(X)$ function 
for $\{m_{g}=0.4H_{0},\, \omega=0.01,\, \alpha_{3}=0.75,\, \alpha_{4}=0.345\}$. 
The red points represent the values of $X_0$ and $G_{2}(X_{0})$ where the initial conditions are satified.
From eq. (\ref{int_const}) we obtain $C=G_{2}(X_{0})$. 
Finally, notice that $G_{2}$ is unbounded as $X\to\infty$;
this implies that for our model the correct past evolution of the background 
is allowed only if $X_{0}>X_{+}$. 

The dark energy equation-of-state parameter $w_{DE, 0}$ is constrained by observations. 
To compute $w$ for the QDMG model we first define the total effective equation-of-state parameter
\be
w_{eff}=-1-\frac{2}{3}\frac{\dot{H}}{H^{2}},
\label{weff}
\ee
and consequently
\be
w_{DE}=\frac{w_{eff}-w_{m}\Omega_{m}-w_{r}\Omega_{r}}{\Omega_{\Lambda}+\Omega_{\sigma}}\, .
\label{wDE}
\ee
We must require that $-1.2<w_{DE, 0}<-0.9$ in agreement with the current limits \cite{2015arXiv150201590P}.

The quasidilaton massive gravity model shows a particular feature -- $\Omega_{DE}$ scales as matter at early times \cite{2013PhRvD..87l3536G}. Indeed, from the analysis above we have $\alpha_{4}>0$ and $C>0$. At early times $G_{2} \sim \alpha_{4}X^{4} = C/a^{4}$. Therefore we find \footnote{To be consistent with observations, $H^{2} \sim a^{-3}$ or $H^{2} \sim a^{-4}$ in the past, so $\Omega_{\sigma} H^{2} \sim a^{-1}$ or $\Omega_{\sigma} H^{2} \sim a^{-2}$.}
\bea
\Omega_{DE} H^{2}  \simeq \Omega_{\Lambda} H^{2}  \simeq m_{g}^{2} \left(\frac{C}{\alpha_4}\right)^{3/4} \frac{\alpha_{3}+\alpha_{4}}{3}\,\,a^{-3}\, .
\label{PastODE}
\eea
It follows that, for redshifts $z \gtrsim 10$, $\Omega_{DE}$ would  contribute to the effective matter energy density! Therefore $\Omega_{DE}$ should be negligible in the radiation era in order to have a viable expansion history.  We demand that $\Omega_{DE}(z_{eq})<0.01$.
\begin{figure}[t]
\begin{center}
\includegraphics[width=0.5\textwidth]{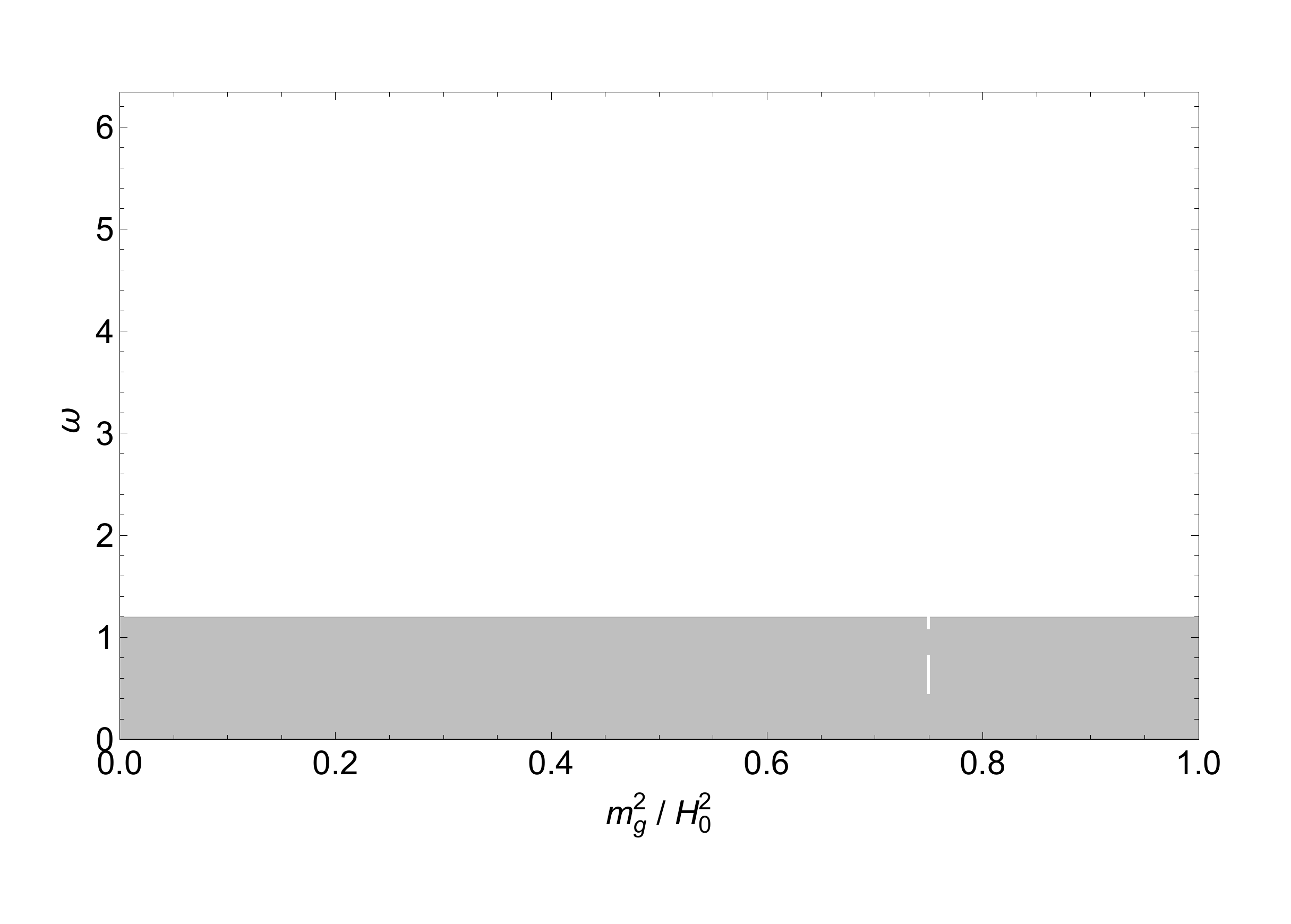}
\caption{(REG1): $m_{g}^{2}/H_{0}^{2} - \omega$ constraints after marginalizing over $\alpha_{3} - \alpha_{4}$.}
\label{MargMgLowO}
\end{center}
\end{figure}

\begin{figure}[t]
\begin{center}
\includegraphics[width=0.5\textwidth]{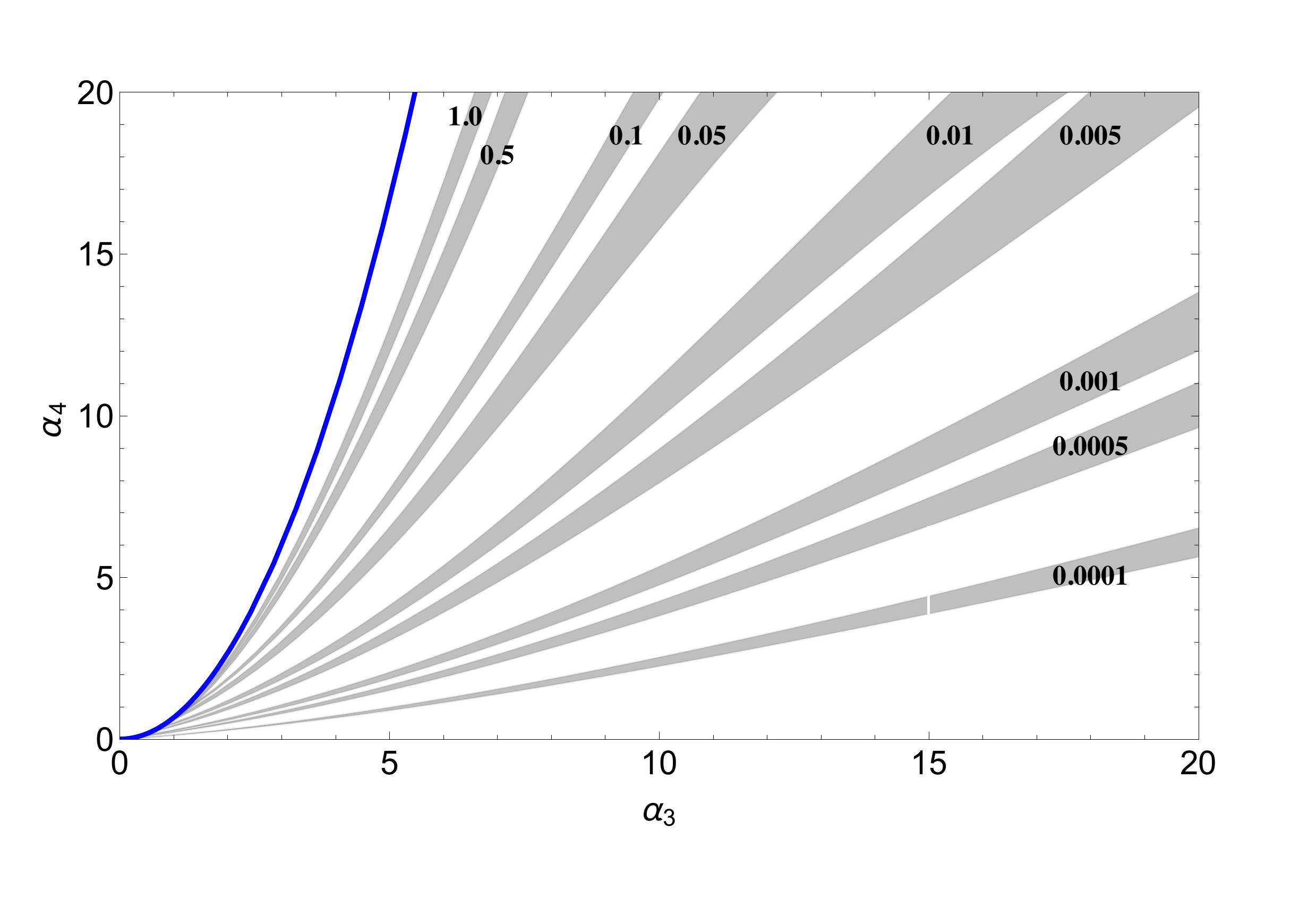}
\caption{(REG1): $\alpha_{3} - \alpha_{4}$ constraints for different $m_{g}^{2}/H_{0}^{2}$ values $(black\,\,bold\,\,numbers)$. We marginalized over $\omega$. The blue line corresponds to the boundary of the region $\alpha_{4}<2\alpha_{3}^{2}/3$, which is the existence condition we obtained from the fixed point analysis.}
\label{MargAlphaLowO}
\end{center}
\end{figure}

In order to identify the allowed ranges of the four parameters of QDMG, 
the main computational obstacle is to find the solutions of the initial condition, 
namely $\Omega_{DE,0}=0.72$.
In principle $\Omega_{DE,0}=0.72$ could have from $1$ to $9$ allowed solutions
for each value of the QDMG parameters.
However, after enforcing all the observational conditions, 
we find that there is never more than $1$ viable solution. 
We identify two disconnected allowed regions in the $4$-dimensional space of parameters, 
one shows just low-$\omega$  values (hereafter REG1) 
and the other one high-$\omega$ values (hereafter REG2).

\begin{figure}[t]
\begin{center}
\includegraphics[width=0.5\textwidth]{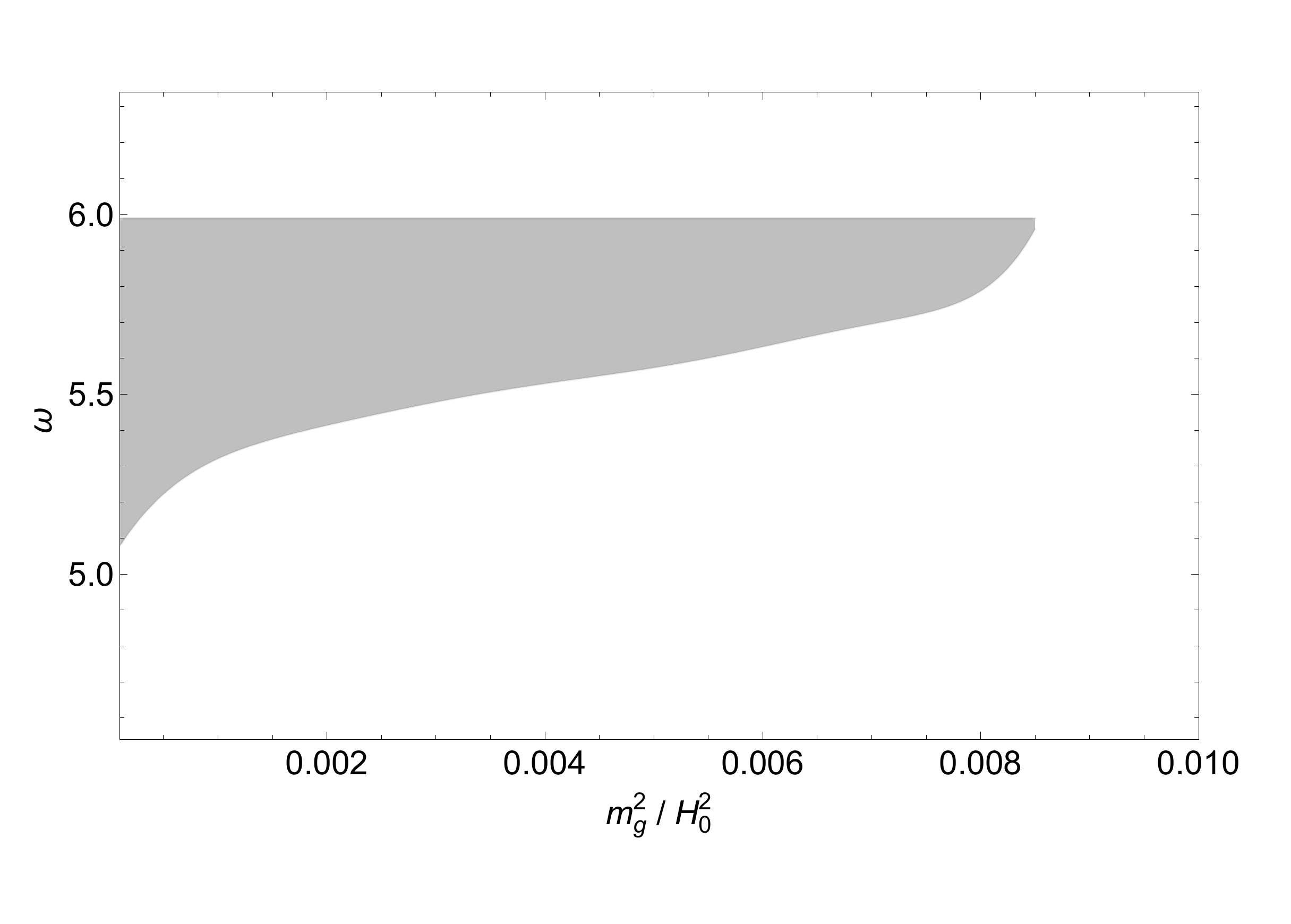}
\caption{(REG2): $m_{g}^{2}/H_{0}^{2} - \omega$ constraints after marginalizing over $\alpha_{3} - \alpha_{4}$.}
\label{MargMgHighO}
\end{center}
\end{figure}

In Fig.~\ref{MargMgLowO} and \ref{MargAlphaLowO} we present the constraints for the (REG1) parameter space. After marginalizing over $\alpha_{3} - \alpha_{4}$ we find that $\omega$ is constrained to $0 < \omega \lesssim 1.2$ as we report in Fig.~\ref{MargMgLowO}.
On the other hand, marginalizing over $\omega$, the contour plot reported in Fig.~\ref{MargAlphaLowO} shows that the $\alpha_{3} - \alpha_{4}$ values are tightly related to the $m_{g}^{2}/H_{0}^{2}$ value $(black\,\,bold\,\,numbers)$. Once we know two of the three $\{m_{g} ,\, \alpha_{3} ,\, \alpha_{4}\}$ parameters, the other one is determined to a good approximation. In other words the quasidilaton massive gravity theory presents a fine-tuning of the parameters. 

\begin{figure*}[t]
\begin{center}
\includegraphics[width=1.0\textwidth]{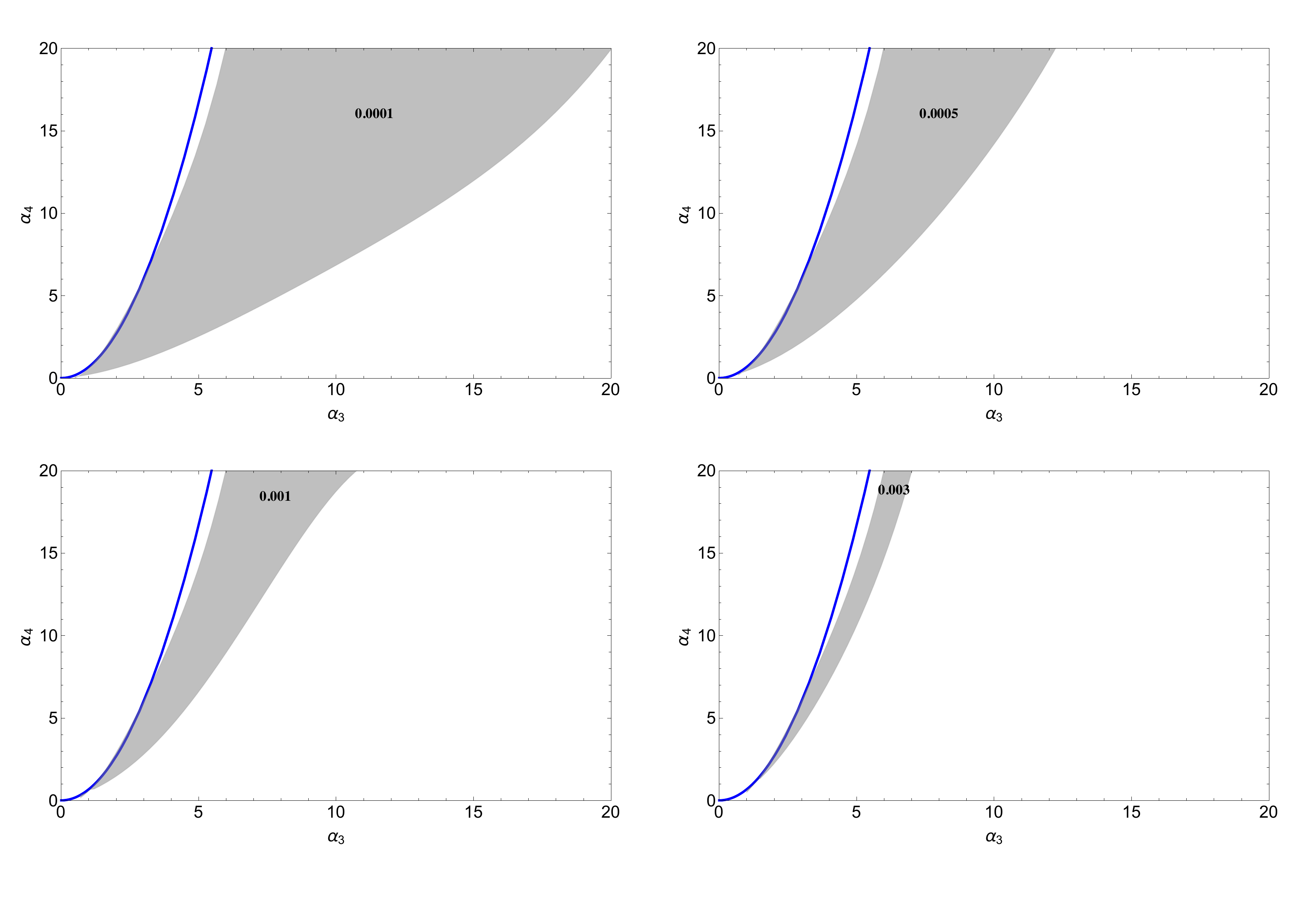}
\caption{(REG2):  $\alpha_{3} - \alpha_{4}$ constraints for different $m_{g}^{2}/H_{0}^{2}$ values $(black\,\,bold\,\,numbers)$. We marginalized over $\omega$. The blue line corresponds to the boundary of the region $\alpha_{4}<2\alpha_{3}^{2}/3$, which is the existence condition we obtained from the  fixed point analysis.}\label{MargAlphaHighO}
\end{center}
\end{figure*}

\begin{figure}[t]
\begin{center}
\includegraphics[width=0.5\textwidth]{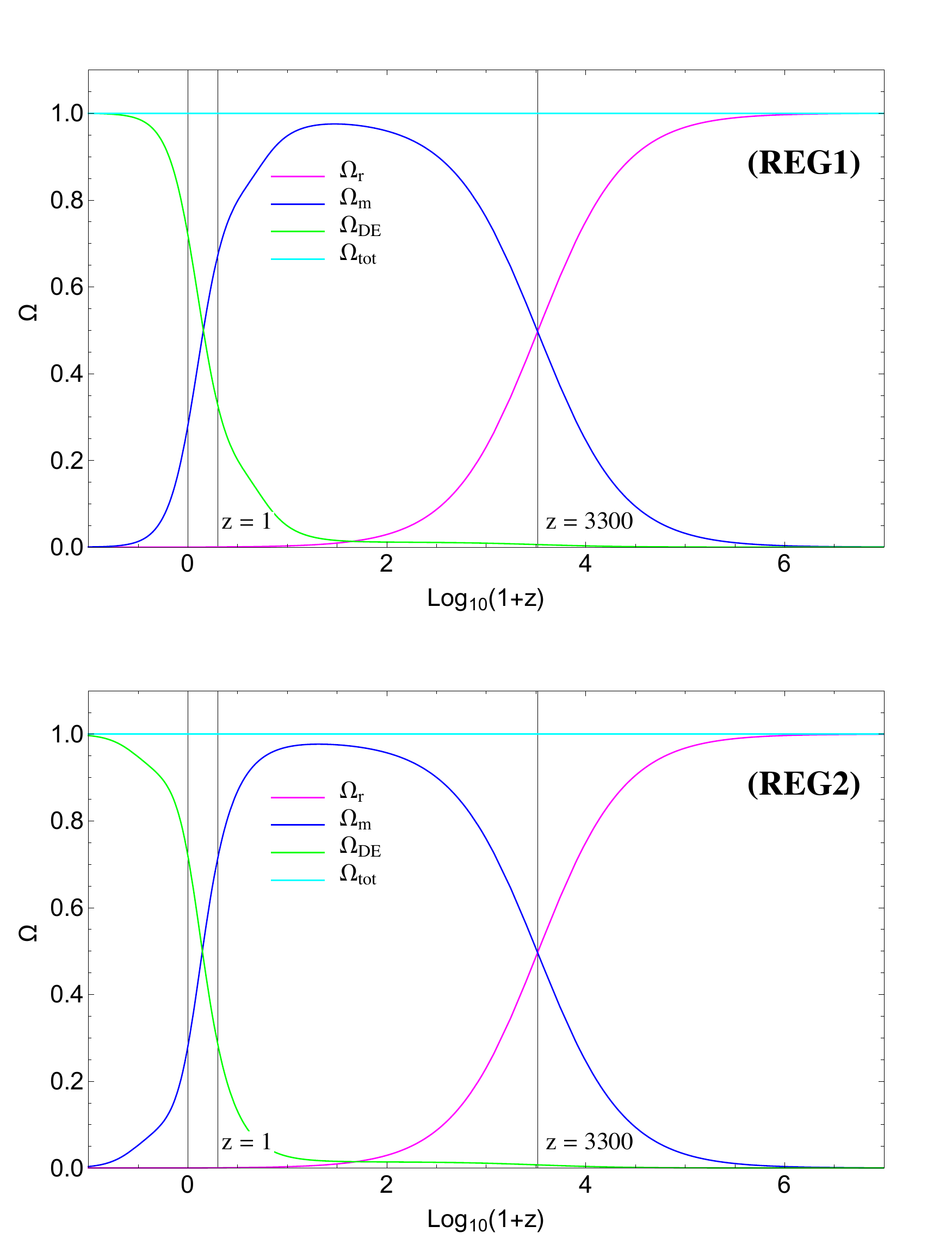}
\caption{{\em {Upper panel}}: expansion history for  $\{m_{g}^{2}/H_{0}^{2}=0.05,\, \omega=0.9484,\, \alpha_{3}=7.059,\, \alpha_{4}=10.63\}$. {\em {Lower panel}}: expansion history for  $\{m_{g}^{2}/H_{0}^{2}=0.0005,\, \omega=5.4211,\, \alpha_{3}=10.797,\, \alpha_{4}=17.680\}$ }
\label{ExpHis}
\end{center}
\end{figure}

Repeating the procedure for (REG2) we find a different behavior. In Fig.~\ref{MargMgHighO} we see that $m_{g}^{2}/H_{0}^{2} \lesssim 0.008$ while the $\omega$ allowed interval depends on $m_{g}$ and it becomes larger as $m_{g}$ decreases.  The $\alpha_{3} - \alpha_{4}$ region is again $m_{g}$ dependent, however the dependence now is different than for (REG1) as we show in Fig.~\ref{MargAlphaHighO}. 
For (REG2) if $m_{g}^{2}/H_{0}^{2} \sim 0.1$ then  $\omega \sim 6$ and $\alpha_{4}$ becomes effectively a function of $\alpha_{3}$, so we find a fine-tuning of $2$ parameters. On the other hand if the graviton mass is small, i.e. $m_{g}^{2}/H_{0}^{2} \lesssim 0.001$, the other parameters are no longer strongly constrained. 

Notice that we find different results than~\cite{2013PhRvD..87l3536G, 2014PhRvD..89h3518B}. In particular they allowed $\omega$ to be negative and they obtain $\omega \lesssim 0.3$.

In our analysis we did not compute the whole expansion history for each point in the four dimensional parameter space for practical computational reasons. As an illustrative example, we choose two set of allowed parameters for (REG1) and (REG2) and we plot the evolution of the energy densities in Fig.~\ref{ExpHis}. As expected, the two panels, are consistent with the observed expansion history.

The parameter fine-tuning we found practically reduces from four to three the effective parameters of the quasidilaton massive gravity theory. We expect that studying the perturbations will further constrain the theory.  Some of those perturbations will be unstable.

\section{\label{GW}Gravity waves}   

In this section we focus on the evolution of the  mass of the gravitational waves.  We consider tensor perturbations around the background   metric solutions,
\begin{equation}
\delta g_{ij}=a^2h^{TT}_{ij}=a^2\int \frac{d^3k}{(2\pi)^{3/2}}h^{TT}_{ij,\vec{k}}\exp(i\vec{k}\cdot\vec{x})+cc,
\end{equation} with $\delta^{ij}h^{TT}_{ij}=0$, and $\partial^jh^{TT}_{ij}=0$. After a straightforward calculation,  one gets  the quadratic Lagrangian for $h^{TT}_{ij,\vec{k}}$ 
\begin{equation}
\mathcal{L}_{GW}=\frac{M_{{\rm Pl}}^{2}}{8}a^3\left[|\dot{h}^{TT}_{ij}|^2-\left(\frac{k^2}{a^2}+M_{GW}^2\right)|h^{TT}_{ij,\vec{k}}|^2\right],
\end{equation}where  the   mass of the gravitational waves $M_{GW}$ is given by \footnote{Note that this expression coincides with the one 
computed for the extended quasidilaton in \cite{2015JCAP...04..010H}
.}
\begin{eqnarray}
M_{GW}^2&=&m_g^2 X\left(3+3\alpha_3+\alpha_4-(1+2\alpha_3+\alpha_4)(1+r)X\right.\nonumber\\
 &+&\left.(\alpha_3+\alpha_4)rX^2\right).
 \end{eqnarray} 
 
We start by computing the ratio $M_{GW}^2/H^2$ at redshift $z=1100$, relevant for CMB, for the two disconnected regions (REG1) and (REG2) defined in the previous section.
An exploration of the values computed reveals there is a minimum and maximum  $M_{GW}^2/H^2$ for each of the regions. The results are presented in 
Table~\ref{MaxMin1100}.
We see that a real mass as large as $M_{GW}\sim 10^{-2} H$ can be obtained, even for our conservative choice  $m_g\leq H_0$. 
For both parameter regions (REG1) and (REG2), we note  the mass can be imaginary. However, the maximum absolute values 
turn out to be much smaller than the Hubble rate, preventing the development of a full  instability. 
It is worth noting that so far signatures  in the Cosmic Microwave Background (CMB) 
due to a non-vanishing  $M_{GW}$ have been studied assuming this mass is always real
\cite{PhysRevD.81.023523, 2012CQGra..29w5026E, 2009JCAP...08..033B}. 
Our results suggest that one should explore also the possibility of having cosmological gravitational 
waves with a small but imaginary mass at the relevant redshifts for CMB.

\begin{figure}[h]
\begin{center}
\includegraphics[width=0.5\textwidth]{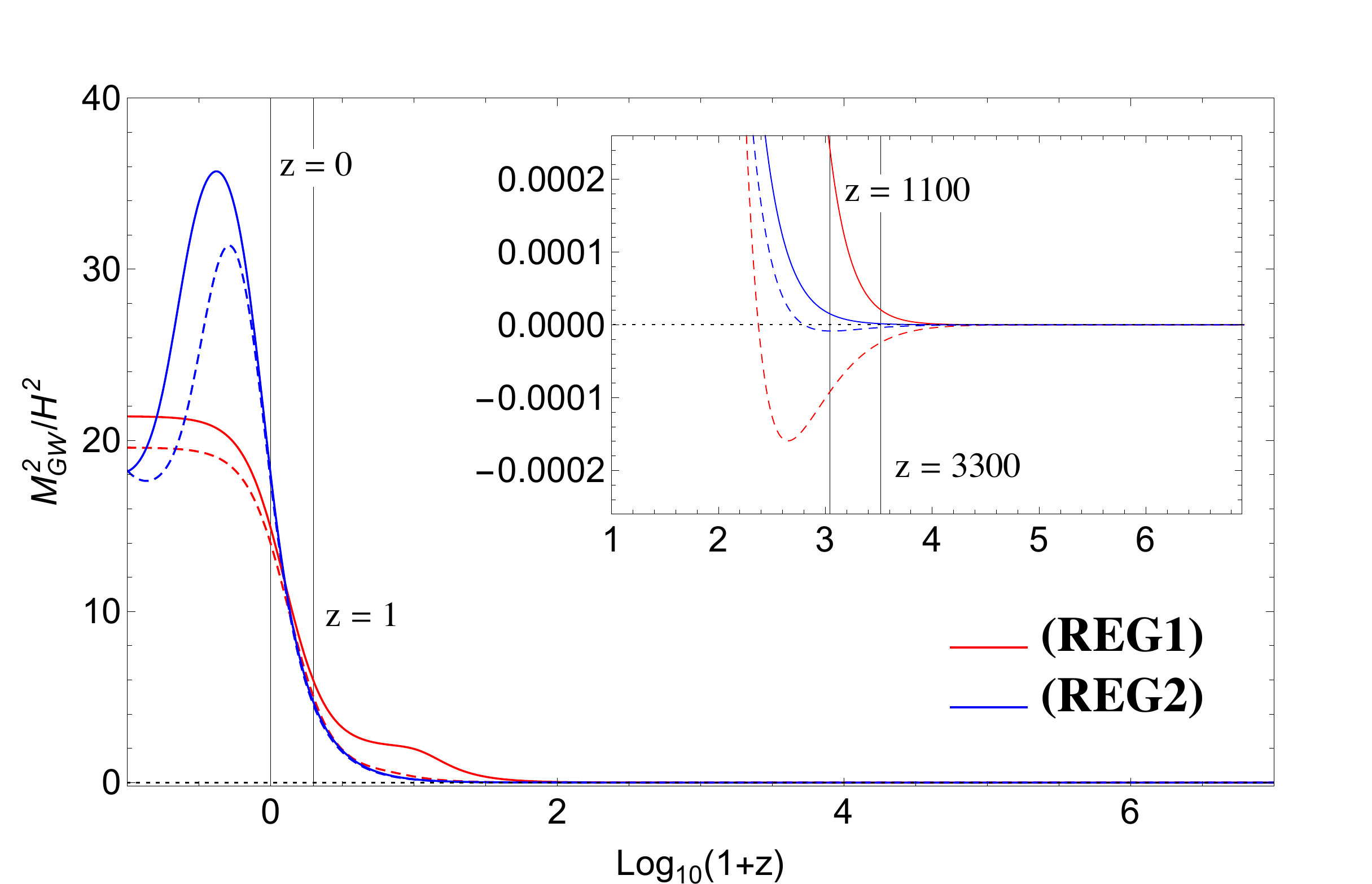}
\caption{Evolution of the mass-squared  of the gravitational waves in units of the Hubble rate for  the  set of parameters given  in Table  \ref{MaxMin1100}. Solid (Dashed) lines correspond to the values of parameters for which we found the maximum (minimum) value  of  $M_{GW}^2/H^2$ at  $z=1100$.  }
\label{MGW}
\end{center}
\end{figure}

In Fig.~\ref{MGW} we plotted the evolution of the ratio $M_{GW}^2/H^2$ 
for the parameters given in Table~\ref{MaxMin1100}.
We  notice that at low  redshifts (and in particular at $z=0$) the mass becomes positive, and is  larger than $H_0$, despite the fact that $m_g\leq H_0$.
In order to assess the generality of this result we computed the ratio $M_{GW}^2/H_0^2$ at  $z=0$ varying the
parameters in the  two disconnected allowed regions of the $4$-dimensional space, and we obtained its maximum and minimum value. The results are shown in Table \ref{MaxMin}. We see that  $M_{GW}$ is  larger than $H_0$, even for values of $m_g\sim10^{-2}H_0$, and 
 it is up to a factor of $\sim 5$ larger than $H_0$ for  $m_g\leq H_0$. 
The existence of a minimum value of $M_{GW}$ that is larger than $H_0$ is remarkable, since this represents a motivated observational threshold. 
That is,  if one could  constrain $M_{GW}$ to be smaller than  $H_0$ one would be able to rule out a self-accelerating explanation of the current acceleration of the universe 
within the QDMG theory.  It would be interesting to see whether an analogous  result  holds for other theories that also aim to provide 
a self-accelerating explanation. Unfortunately, current experiments are still far from probing $M_{GW}\sim H_0$ \cite{Agashe:2014kda}.
Moreover,  the upper limits one can obtain are in general model dependent, since they are based on assumptions involving different scales of the 
theory.  This represents a challenge for both theory and observations, and  highlights the ongoing importance of working out  predictions within the framework of specific models of modified gravity.

\begin{table}[ht]
\begin{center}
\begin{tabular}{ |c|c|c|c|c| }
    \hline
&    \multicolumn{2}{|c|}{(REG1)}&\multicolumn{2}{|c|}{(REG2)}\\
$z=1100$  & \multicolumn{2}{|c|}{Max.$\,\,\,\,\,\,\,\,\,\,\,\,\,\,\,\,{}$ Min.} & \multicolumn{2}{|c|}{Max.$\,\,\,\,\,\,\,\,\,\,\,\,\,\,\,\,{}$ Min.} \\
    \hline
 $ M_{GW}^2/H^2$ & $2.5\times 10^{-4}$ &$ -9.2\times 10^{-5}$&$1.5\times 10^{-5}$& $-8.5\times10^{-6}$\\
  $m_g^2/H_0^2$& $\,\,\,1\,\,\,$ &1& $10^{-4}$ &$8.45\times 10^{-3}$ \\
$ \omega$ & 1.03 &  0.32  &5.99&5.95\\
  $\alpha_3$ & 6.60 &  6.80 &5.83&3.56\\
  $\alpha_4$  & 19.61 & 19.87  &20&8.44\\
    $X_0$ &1.82& 1.85 &1.73&1.94\\
  \hline
\end{tabular}
\end{center}
\caption[crit]{\label{MaxMin1100} Maximum and minimum values of  $M_{GW}^2/H^2$ at  $z=1100$ and the corresponding values of the parameters $\{m_g,\omega,\alpha_3,\alpha_4\}$.  For completeness, the value of $X$ at $z=0$ is also presented. } 
\end{table}

\begin{table}[ht]
\begin{center}
\begin{tabular}{ |c|c|c|c|c| }
    \hline
&    \multicolumn{2}{|c|}{(REG1)}&\multicolumn{2}{|c|}{(REG2)}\\
$z=0$  & \multicolumn{2}{|c|}{Max.$\,\,{}$ Min.} & \multicolumn{2}{|c|}{Max.$\,\,{}$ Min.} \\
    \hline
 $ M_{GW}^2/H_0^2$ & 25 & 2.5&18&5 \\
  $m_g^2/H_0^2$& $\,\,\,1\,\,\,$ & $10^{-4}$ &$10^{-4}$ &$10^{-4}$ \\
$ \omega$ & 0.08 &  0.08  &6&5.05\\
  $\alpha_3$ & 0.86 &  3.46 &5.83&1.70\\
  $\alpha_4$  & 0.48 & 0.55  &20&0.49\\
    $X_0$ &4.67& 19.59  &1.73&10.99\\
  \hline
\end{tabular}
\end{center}
\caption[crit]{\label{MaxMin} Maximum and minimum values of  $M_{GW}^2/H_0^2$ at  $z=0$, the corresponding values of the parameters $\{m_g,\omega,\alpha_3,\alpha_4\}$,  and the value of $X$ at $z=0$. }
\end{table}

\section{\label{}Conclusions}   

The combination of General Relativity and the Standard Model of particle physics are demonstrably 
and remarkably successful descriptions of the world on scales up to and including the solar system.   
On larger scales, there is a need either to modify the theory of gravity
or to introduce new forms of dark matter and dark energy.  
The most parsimonious solution would be to identify candidates for the latter in the Standard Model, 
and such candidates may exist for dark matter (see for example \cite{1984PhRvD..30..272W,1990NuPhB.345..186L}) and
evade existing constraints \cite{2015MNRAS.450.3418J}, although the phenomenological successes of 
MOND (see for example \cite{2015CaJPh..93..250M}) cannot be entirely dismissed as an indication of the need to modify gravity on galactic scales.
For the observed cosmic acceleration, the situation is even less clear. A cosmological constant is the
 canonical explanation, but despite decades of attempts has as yet no clear explanation in the Standard Model.  The need for observational probes of possible dark energy and modified gravity explanations
 is thus paramount.

 One possibility would be to develop some general phenomenological classification of possible deviations
 of gravity from GR. The Parametrized post-Newtonian approach is one such program, in the context of 
 almost-Schwarzschild backgrounds. Such generic approaches have also been attempted in the 
 cosmological context (eg.   \cite{2004PhRvD..69d4005L,2007PhRvD..76j4043H}).   However, in the context of a highly non-linear theory such as GR, the observational consequences of small theoretical departures from GR can be quite ideosyncratic. While phenomenological parametrization of observables may be convenient, and even
 useful, they may not capture (or may capture poorly) the specific phenomena or behaviour that result
 from actual models.  Careful examination of specific individual models can therefore be both instructive and essential.

In this paper, we have studied the (homogeneous) cosmological solutions of quasidilaton massive gravity.
A study of the linear perturbations around the asymptotic self-accelerated cosmological solution 
of this theory (which corresponds to a De Sitter background metric) has been done in \cite{2013CQGra..30r4005D, 2013arXiv1304.0449E}.
These studies have revealed that the kinetic term of one of the scalar perturbations 
becomes negative for short wavelengths, indicating that the theory may have a ghost instability 
that shows up at short distances. This is indeed the case at linear level. 
Several authors \cite{2014PhLB..728..622D, 2013PhRvD..88l4006D} 
have therefore extended the theory by allowing for a new coupling, 
which can be  properly adjusted to make the scalar sector stable at linear level. 
This extended quasidilaton massive gravity theory (EQDMG), 
has been considered by other authors \cite{2014PhRvD..90j4008M, 2015PhRvD..91d1301K, 2015JCAP...04..010H}. 

Although this current reconsideration of the background cosmological solutions of QDMG was
performed  as a first step for a full analysis of the EQDMG, it revealed important attributes
of the QDMG cosmology, which we expect to carry over qualitatively or in detail to EQDMG.
The first is that observationally viable QDMG cosmologies require fine-tuning of parameters. 
In particular, the allowed values of the graviton mass parameter, $m_g$ is a tightly constrained 
function of the coupling constants $\alpha_3$ and $\alpha_4$, 
with only a very narrow tolerance around a central value $m_g(\alpha_3,\alpha_4)$.
This fine-tuning, and the  precise value of $m_g(\alpha_3,\alpha_4)$, is dictated by observational constraints on the dark energy properties.

The second observation is that some small (but possibly non-negligible) fraction 
of what manifests as $\Omega_{DE}$ (i.e. $p/\rho\simeq-1$) today, was $\Omega_{m}$ 
(i.e. $p/\rho\simeq0$) in the past.  The transition from one equation of state to the other
was sudden and probably not well-captured by a linear parametrization of $w(z)$. The expected
difference between $\Omega_m$ at high redshift (as measured in the CMB) and at low redshift
(as measured, say in large scale structure), could be the source of recently noted tensions
in different determinations of $\Omega_m$ \cite{2015arXiv150201597P,2015A&A...574A..59D}. While the details of these behaviors
of the background cosmomology are likely to be altered in EQDMG, it is plausible that
these qualitative features are robust.

We have also analyzed the phenomenology of the graviton.   
The governing equations for the graviton mass $M_{GW}$ 
(which is not equal to the graviton mass parameter $m_g$)
are the same in QDMG and EQDMG. 
We therefore expect to gain useful insights for the extended model
provided that the background solutions do not depend sensitively on the new parameter $\alpha_\sigma$ of the extended theory.  
We find that the graviton mass-squared  typically is negative at redshifts well above $z=1$, 
indicating an instability.  This includes redshifts $z\simeq 10^3-10^4$ where such physics
may well imprint itself on the CMB.   At any given time $\vert M_{GW}^2 \vert \ll H^2$ , 
so we do not expect the  instability to lead to  many e-foldings of growth.
Nevertheless, if this persists in EQDMG, it may be another opportunity to 
see evidence of modified gravity in CMB observations. 

Regarding vector perturbations, according to equations (4.16) and  (4.17) of \cite{2015JCAP...04..010H}, the square of the speed of propagation, $c_V^2$, can be recast as
 $c_V^2= \kappa_V / M_{GW}^2$, where the absence of ghost instability is guaranteed provided $\kappa_V>0$ \footnote{We note this inequality is satisfed in the QDMG case for parameters in the allowed regions}.  Notice in particular that when $M_{GW}^2$ becomes negative,  the absence of ghost instability implies  $c_V^2$ becomes also negative.  Therefore, we expect that a detail analysis of the perturbations will further reduce the region of allowed parameters.  
 
In a future work, we will therefore extend our analysis to the EQDMG theory, taking into account the constrains from the study of the perturbations, anticipating hopefully that these observable effects will indeed persist.

\begin{acknowledgments} 
We thank C.~de Rham for helpful comments, and G.~Gabadadze, M.~Fasiello, D.~M{\"u}ller and A.~Tolley for discussions. DLN is thankful the Case Western Reserve University for hospitality when this work was initiated. We acknowledge the use of the xAct - xPand package for Mathematica \cite{2013CQGra..30p5002P, xAct}. SA and GDS are supported by a Department of Energy grant DE-SC0009946 to the particle astrophysics theory group at CWRU.
\end{acknowledgments}

\bibliography{MyBib2}

\end{document}